\documentclass[a4paper,12pt]{article}  
\usepackage[dvips]{graphicx}
\usepackage{amssymb}
\usepackage{cite}
\newcommand{\goes}{\rightarrow} 
 
\newcommand{\GeV}{\; \mathrm{GeV}} 
\newcommand{\TeV}{\; \mathrm{TeV}} 
\newcommand{\lapproxeq}{\lower .7ex\hbox{$\;\stackrel{\textstyle  
<}{\sim}\;$}} 
\newcommand{\gapproxeq}{\lower .7ex\hbox{$\;\stackrel{\textstyle  
>}{\sim}\;$}} 
\newcommand{\stackdown}[2]{\lower 1.4ex\hbox{$\;\stackrel{\textstyle{#1}}  
{\scriptstyle{#2}}\;$}}
\newcommand{\beq}{\begin{equation}} 
\newcommand{\eeq}{\end{equation}} 
\newcommand{\bea}{\begin{eqnarray}} 
\newcommand{\eea}{\end{eqnarray}}
\newcommand{\lsp}{\tilde{\chi}}

\newcommand{\relic}{\Omega_{\lsp}\,h_0^2} 
 
\newcommand{\etal}{\textit{et. al.}}

\makeatletter 
\def\slash{\@ifnextchar[{\fmsl@sh}{\fmsl@sh[0mu]}} 
\def\fmsl@sh[#1]#2{%
  \mathchoice 
    {\@fmsl@sh\displaystyle{#1}{#2}}%
    {\@fmsl@sh\textstyle{#1}{#2}}%
    {\@fmsl@sh\scriptstyle{#1}{#2}}%
    {\@fmsl@sh\scriptscriptstyle{#1}{#2}}} 
\def\@fmsl@sh#1#2#3{\m@th\ooalign{$\hfil#1\mkern#2/\hfil$\crcr$#1#3$}} 
\makeatother 

\begin{document} 
\begin{titlepage} 
 
\begin{flushright} 
\parbox{4.6cm}{HEPHY-PUB 739/01\\
               UA/NPPS-06-01\\
               hep-ph/0106345} 
\end{flushright} 
\begin{centering} 
\vspace*{1.5cm} 
 
{\large{\textbf {Implications of the 
Pseudo-Scalar Higgs Boson \\ 
in determining the Neutralino Dark Matter
}}}\\
\vspace{1.4cm} 
 
{\bf A.~B.\ Lahanas} $^{1}$ \, and \, {\bf V.~C.~Spanos} $^{2}$  \\ 
\vspace{.8cm} 
$^{1}$ {\it University of Athens, Physics Department,  
Nuclear and Particle Physics Section,\\  
GR--15771  Athens, Greece}\\ 
 
\vspace{.5cm} 
$^{2}$ {\it  Institut f\"ur Hochenergiephysik der \"Osterreichischen Akademie
der Wissenschaften,\\
A--1050 Vienna, Austria}\\ 
\end{centering} 
\vspace{3.cm} 
\begin{abstract}
In the framework of the Constrained Minimal Supersymmetric Standard Model 
(CMSSM) we discuss the impact of the
pseudo-scalar Higgs boson in delineating regions of the
parameters which are consistent with cosmological data and E821 data 
on the anomalous magnetic moment of muon.
For large values of the parameter $\tan \beta > 50$,
cosmologically allowed corridors of large
$m_0$, $M_{1/2}$ are opened, 
due to the $s$-channel pseudo-scalar exchange in the pair annihilation of the
lightest of the neutralinos to $b \bar{b}$ or $\tau \bar{\tau}$, 
which dominates in this region. However, 
no such corridors are found for values  $\tan \beta < 50 $. Combining 
cosmological and E821 data puts severe upper limits on sparticle masses. 
We find that at LHC, but
even at a $e^{+}e^{-}$ linear collider with center of mass energy
$\sqrt{s} = 800\GeV$, such as TESLA,   
supersymmetry can be discovered, if it is based on the CMSSM.
  
\end{abstract} 
\end{titlepage} 
\newpage 
\baselineskip=20pt 

\section{Introduction}
In the framework of supersymmetric models with $R$-parity conservation,
it has been argued that for large $\tan\beta$
the neutralino relic density ($\relic$) can be
compatible with the recent cosmological data which favour
small values for $\relic$. In this regime 
the neutralino ($\lsp$) pair annihilation 
through $s$-channel
pseudo-scalar Higgs boson ($A$) 
exchange leads to an enhanced annihilation cross sections
reducing significantly the relic 
density \cite{Drees}, 
while the heavy $CP$-even Higgs ($H$) exchange
is $P$-wave suppressed and not that important.
The importance of this mechanism, in conjunction with the recent
cosmological data which favour small values of the Dark Matter (DM)
relic density,
has been stressed in \cite{LNS}. The same mechanism has been
also invoked in Ref.~\cite{focus} 
and \cite{Ellis} where it has been shown that it enlarges the cosmologically
allowed regions. In fact 
Cosmology does not put severe upper
bounds on sparticle masses, and soft masses can be in the TeV region,
pushing up the sparticle mass spectrum to regions that might escape detection
 in future planned accelerators. 
Such upper bounds are imposed, however, by
the recent $g-2$ E281 data \cite{E821} 
constraining the CMSSM in such a way that
supersymmetry will be accessible to LHC or other planned $e^{+}e^{-}$ linear
colliders if their center of mass energy is larger than 
about $1.2\TeV$ \cite{ENO}.
The bounds put by the $g-2$ has been the subject of intense phenomenological 
study \cite{ENO,g-2,Leszek,Narison}, and although the
situation has not been definitely settled, supersymmetry emerges as
a prominent candidate
in explaining the discrepancy between the Standard Model predictions and
experimental measurements.

In this study we undertake the problem of calculating the neutralino relic
density, in the framework of the CMSSM, paying special attention
to the pseudo-scalar Higgs exchange which dominates in the large $\tan \beta$
region.
In this regime the cosmologically allowed domains depend sensitively on this
mechanism and in conjunction with the bounds put by the $g-2$ measurements can 
severely constrain the CMSSM predictions. In particular, we find that
cosmologically allowed corridors of large $m_0, M_{1/2}$ values open up
for $\tan \beta > 50 $, which however have little overlap with the
regions allowed by the E821 data. The constraints imposed on the sparticle
spectrum and the potential of discovering CMSSM in future accelerators
are discussed.

\section{The role of the pseudo-scalar Higgs boson mass}
The $\lsp \lsp$ fusion to the pseudo-scalar Higgs boson, $A$, which
subsequently decays to a $b \bar{b}$ or a $\tau \bar{\tau}$, becomes the
dominant annihilation mechanism for large $\tan \beta$, when the pseudo-scalar
mass $m_A$ approaches twice the neutralino mass,  
$m_A \simeq 2 m_{\lsp}$.
In fact by increasing $\tan \beta$ the mass $m_A$ decreases, while the
neutralino mass remains almost constant, if the other parameters are kept
fixed. Thus  $m_A$ is expected eventually to enter into the regime in which
it is close to the pole value $m_A\,=\, 2 m_{\lsp}$, and the
pseudo-scalar  Higgs exchange dominates. 
It is interesting to point out that in a previous analysis of the direct
DM searches \cite{LNSd}, we had stressed 
that  the contribution of the $CP$-even Higgs bosons exchange
to the LSP-nucleon scattering cross sections increases with $\tan \beta$.
Therefore in the large $\tan \beta$ 
regime one obtains the highest possible rates
for the direct DM searches. Similar results are presented in Ref. 
\cite{Kim}.

In the framework of the CMSSM
the chargino mass bound as well as the recent LEP Higgs mass bound \cite{LEP} 
already
exclude regions in which $\lsp$ has a large Higgsino component, and thus 
in the regions of interest the $\lsp$ is mainly a bino.  A bino is 
characterized by a very small coupling to the pseudo-scalar Higgs $A$, 
however the largeness of
$\tan \beta$ balances the smallness of its coupling giving a sizeable
effect when $m_A\;\simeq \; 2 m_{\lsp}$, making the $s$-channel
pseudo-scalar exchange important.

It becomes obvious from the previous discussion that an unambiguous and
reliable determination of the $A$-mass, $m_A$, 
is demanded before one embarks on
to calculate the neutralino relic density especially 
in regions where the $s$-channel
pseudo-scalar exchange dominates. In the constrained SUSY models,
such as the CMSSM, 
$m_A$ is not a free parameter but 
is determined once $m_0$, $M_{1/2}$, $A$ as
well as $\tan \beta$ and the sign of $\mu$, $sign(\mu)$, are given.
$m_A$  depends sensitively on the Higgs
mixing parameter, $m_3^2$, which is determined from minimizing 
the  one-loop corrected effective potential. A subtlety arises  
for large $\tan \beta$ values since
the corrections are relatively large due mainly to the smallness of the
Higgs mixing parameter. In order to handle this 
we calculate the effective potential using as
reference scale the average stop scale
$Q_{\tilde t}\simeq\sqrt{m_{{\tilde t}_1} m_{{\tilde t}_2} }$ \cite{scale}.
At this scale the contributions of the third generation sfermions are
small.
However, other contributions may not be negligible at this scale and should
be properly taken into account. 
In particular, the neutralino and chargino contributions to the effective
potential 
should be included for a more reliable calculation. These do not vanish
at $Q_{\tilde t}$ since their masses, determined by the 
gaugino masses $M_1, M_2$ and the $\mu$ value, may be quite different
from $Q_{\tilde t}$.
Their inclusion to the effective potential improves the mass of
the $A$-Higgs, as this is calculated from the effective potential, yielding 
a result that is scale independent and 
approximates the pole mass to better than $2 \%$
if we also include the scale dependent logarithmic contributions from the
wave function renormalization 
$\Pi(0) \;-\;\Pi(m_A^2)$ 
 \cite{KLNS}.
I the present work and
for a more accurate determination of the pseudo-scalar Higgs we use the pole
mass using the expression of Ref. \cite{BMPZ}.

A more significant correction, which 
drastically affects the pseudo-scalar mass 
arises from the gluino--sbottom and chargino--stop corrections to the bottom
quark Yukawa coupling \cite{mbcor,wagner,BMPZ,arno,eberl,car2}.
Taking these effects into account the tree-level relation between
the bottom mass and the corresponding Yukawa coupling is
modified according to 
\begin{eqnarray}
m_b\;=\;v_1 \,(\;h_b + \Delta h_b \tan \beta\;)\,,  \label{bottom}
\end{eqnarray}
where $\Delta h_b$ is 
\begin{eqnarray}
\frac{\Delta h_b}{h_b}\;=\;\frac{2 \alpha_s}{3 \pi}
m_{\tilde g} \mu  \;G(m^2_{{\tilde b}_1}, m^2_{{\tilde b}_2}, m_{\tilde g}^2)
\;-\;\frac{h_t^2}{16 \pi^2} \; \mu A_t \;
G(m^2_{{\tilde t}_1}, m^2_{{\tilde t}_2}, {\mu}^2)     \;.
\label{arnow}
\end{eqnarray}
In this the first (second) term is the gluino, sbottom (chargino, stop)
corrections. In the second term we have ignored the small
electroweak mixing effects.
The function $G$ in Eq.~(\ref{arnow})
is the one used in Refs.~\cite{arno,wagner,car2} which  
in \cite{wagner} is denoted by $I$.
It is known that the proper resummation of these corrections
is important for a correct determination of $h_b$ \cite{eberl,car2}.
Eqs.~(\ref{bottom}) and (\ref{arnow}) agree with those  of Ref.~\cite{car2}
and therefore these corrections have been properly resummed
{\footnote{
Our $\mu$ and the soft gaugino masses  differ in sign from those of
Ref.~\cite{wagner,car2}, while $A_t$ has the same sign.}}.

\begin{figure}[t] 
\centering\includegraphics[height=10cm,width=10.5cm]{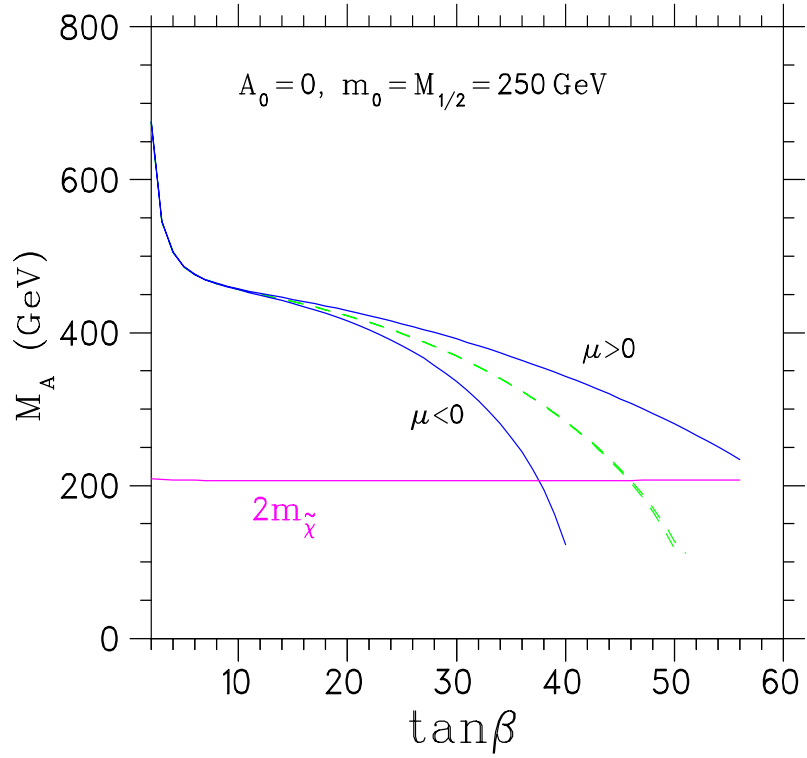}
\caption[]{The pseudo-scalar Higgs masses for $\mu>0$ and $\mu<0$ as function 
of $\tan \beta$ (solid lines). The dashed lines are the same masses when
the supersymmetric corrections to the bottom Yukawa
coupling are ignored. }
\label{fig1}  
\end{figure}

These are very important, especially the
SQCD corrections, and should be duly taken into account.
The important point is that these affect differently the $\mu>0$ and $\mu<0$
values of $m_A$ in a drastic way.
In figure \ref{fig1} we show the behaviour for the
pseudo-scalar Higgs boson mass as function 
of the parameter $\tan \beta$ for the inputs
shown in the figure,  
for both signs of $\mu$, with and without the
aforementioned corrections.
Also plotted is the
value of the quantity $2\;m_{\lsp}$. One observes that in the absence of these
correction to the bottom Yukawa coupling, 
the pseudo-scalar Higgs mass in the two
cases differ little and meet the $2\;m_{\lsp}$ line at $\tan \beta \simeq 40$.
Obviously the $\tan \beta$ region around this value leads to enhanced
neutralino annihilation cross sections through $A$-exchange since we are
in the vicinity of a pole. However, when the aforementioned corrections are
taken into account the values of the pseudo-scalar mass in the two cases split
as shown in the figure. That corresponding to the $\mu>0$  is moving upwards
and that to $\mu<0$ downwards. Thus only the second can reach the
pole value $2\;m_{\lsp}$ at a smaller $\tan \beta$ however. The mass
corresponding to $\mu>0$ stays away from this, never reaching it, at least
in the case shown, since for higher values of $\tan \beta$ we enter regions
which are theoretically forbidden.
Actually in these regions Electroweak
Symmetry Breaking does not occur.
The $\mu<0$ case does not stay comfortably well with the
$b \goes s + \gamma$ process, as well as with the observed discrepancy
of the $g-2$ data, if the latter are attributed to supersymmetry, and  
therefore we shall discard it in the sequel.

It is obvious from the previous discussion that the crucial parameter in
this analysis is the ratio ${m_A}/{2 m_{\lsp}}$. The calculation of the
pseudo-scalar Higgs boson mass $m_A$ we have discussed in detail before.
For the calculation of the
$\lsp$ mass we use the one-loop corrections of Ref.~\cite{BMPZ}.
These result to corrections as large as $5\%$, in some cases, reducing the
aforementioned ratio, making it to be closer to the pole value.

\section{Numerical results--Discussion}
Before embarking on to present our results
we shall comment on our numerical
analysis employed in this paper. This will be useful when comparing the
results of this paper with those of other works, which  use different 
numerical schemes in determining the mass parameters of the CMSSM. The
predictions for the sparticle spectrum, including the mass of the
pseudo-scalar Higgs, as well as the calculation 
of the relic density itself, may be
sensitive to some of the parameters and the particular scheme employed.

In our analysis we use two-loop renormalization group equations (RGE), in
the $\overline{DR}$ scheme, for all masses and couplings involved. The
unification scale $M_{GUT}$ is defined as the point at which the gauge
couplings
${\hat \alpha}_1$ and ${\hat \alpha}_2$ meet but we do not enforce
unification of the
strong coupling constant ${\hat \alpha}_3$ with ${\hat \alpha}_{1,2}$ 
at $M_{GUT}$. The experimental value of the $\overline{MS}$ 
strong coupling constant at $M_Z$ is an input in
our scheme and this is related to ${\hat \alpha}_3$ through
$\alpha_s(M_Z)={\hat \alpha}_3(M_Z) / (1 - \Delta {\hat \alpha}_3 )$,
where
$\Delta {\hat \alpha}_3$ are the threshold corrections. Enforcing unification
of ${\hat \alpha}_3$ with the rest of the gauge couplings usually results in
values 
for $\alpha_s(M_Z)$ larger than the experimental values, if the two-loop RGE's
are used.
For this reason we abandon gauge coupling unification.
For the determination of the
gauge couplings ${\hat \alpha}_{1,2}$ we use as inputs the
electromagnetic coupling constant $\alpha_0$, 
the value of the Fermi coupling constant $G_F$,
and the $Z$-boson mass $M_Z$. From these we determine the
weak mixing angle through the relation
${\hat s}^2 {\hat c}^2 ={\pi \alpha_0}\;/\;{\sqrt{2} M_Z^2 G_F}
{(1-\Delta {\hat r}})$. 
The value of the electromagnetic coupling constant at $M_Z$ in the
$\overline{DR}$ scheme is calculated through
${\hat \alpha}(M_Z) = \alpha_0 /(\;1 - \Delta {\hat \alpha}_{em}\;)$, where
$\;\Delta {\hat \alpha}_{em}\;$ are the appropriate threshold
corrections (for details see Ref.~\cite{BMPZ}).
In each iteration ${\hat s}^2$ and ${\hat \alpha}$ are extracted and from
these the values of ${\hat \alpha}_{1,2}$ at $M_Z$ are determined.
In the equations above
all hatted quantities are 
meant to be in the $\overline{DR}$ scheme.   

Our remaining inputs, in running the RGE's, 
are as usual the soft SUSY breaking parameters $m_0$, $M_{1/2}$, $A_0$,
$\tan \beta$ and the sign of the parameter $\mu$. The top and tau
physical masses,  
$M_t, M_{\tau}$, as well as the $\; \overline{MS}$ bottom
running mass $m_b(m_b)$ are also inputs.
The default values for the aforementioned masses are
$M_t = 175 \GeV, M_{\tau} = 1.777\GeV$ and $m_b(m_b) = 4.25\GeV$. 
For the determination of the bottom and tau running masses, and hence
their corresponding Yukawa couplings,
at the scale $M_Z$, we run $S{U_c}(3) \times U_{em}(1)$ RGE's 
using three-loop RGE's for the strong coupling constant.
We also include two-loop contributions in the electromagnetic coupling
and mixings of the electromagnetic with the strong coupling constant.
The latter is as important as the three-loop strong
coupling constant contribution to the RGE's. At the end of the
running $ \overline{MS}$ masses are converted to
 $ \overline{DR}$ in the usual way. This determines the bottom and tau 
Yukawa couplings at $M_Z$. We recall that the corrections to the bottom
Yukawa coupling of Eq.~(\ref{arnow}) should be duly taken into account.
For the determination of the top Yukawa coupling at $M_t$ 
we take into account all dominant corrections relating the pole 
to its running mass. By running
the RGE's we can have the top Yukawa coupling at $M_Z$.
Thus our analysis resembles that followed in Ref.~\cite{BMPZ}.

For the determination of the Higgs and Higgsino mixing 
parameters, $m_3^2$ and  $\mu$, 
we solve the minimization conditions with the one-loop corrected 
effective potential in which all particle contributions are taken into
account.
The minimization is performed using as reference scale the average stop scale
$Q_{\tilde t}\simeq\sqrt{m_{{\tilde t}_1} m_{{\tilde t}_2} }$.
Thus in each
run we determine $m_3^2(Q_{\tilde t}), \mu(Q_{\tilde t})$.

Regarding the calculation of the lightest supersymmetric particle (LSP) 
relic abundance, the Boltzmann equation 
is solved numerically using the machinery
outlined in Ref.~\cite{LNS}. The coannihilation effects  
in regions where the right-handed stau, 
${\tilde \tau}_R$, approaches in mass the LSP,
are properly taken into account.
Caution should be taken in regions where the ratio
${m_A}/{2 m_{\lsp}}$ is close to unity. 
This signals the vicinity of
a pole in which case the traditional non-relativistic expansion
breaks down. On the pole the annihilation cross section 
through the pseudo-scalar Higgs $s$-channel exchange is large
and its width is important in
determining its size. The rescaled pseudo-scalar Higgs boson width is
$\Gamma_A /m_A \simeq 10^{-2}$, that is resembles that of the $Z$-boson,
and hence for values of the ratio ${m_A}/{2 m_{\lsp}}$
larger than about 1.2 (see \cite{griest}) we are away from the pole region.  
We have found that in the $m_0, M_{1/2}$ plane, and for both parameters less
than $1\TeV$, this ratio approaches unity   
only for very large values of $\tan \beta > 50$. For such values
of $\tan \beta $ the pseudo-scalar Higgs dominates the $\lsp$
pair annihilation, leading to large cross sections and therefore 
cosmologically acceptable relic densities.
Thus cosmologically allowed $m_0, M_{1/2}$ corridors open up for
$\tan \beta > 50$ which were absent for lower values of $\tan \beta$.
These are the same corridors observed in the analysis of Ref.~\cite{Ellis}
which however show up for lower values of $\tan \beta$.

\begin{figure}[t] 
\includegraphics[height=7.5cm,width=7.5cm]{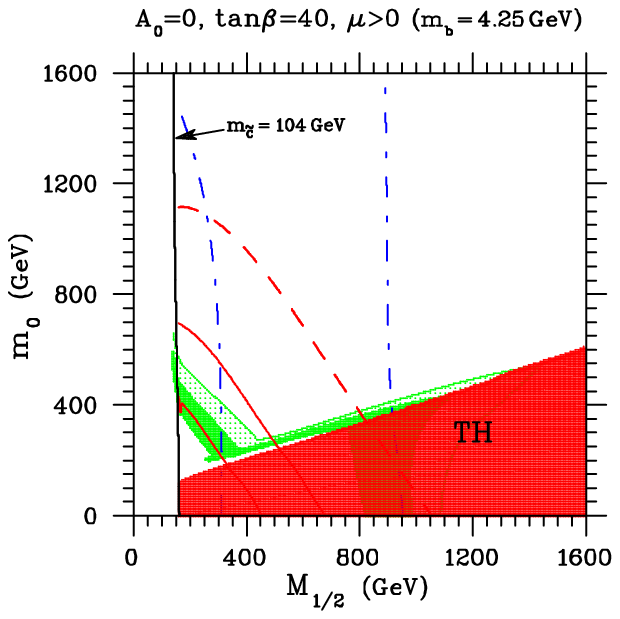}
\hspace*{.5cm}
\includegraphics[height=7.5cm,width=7.5cm]{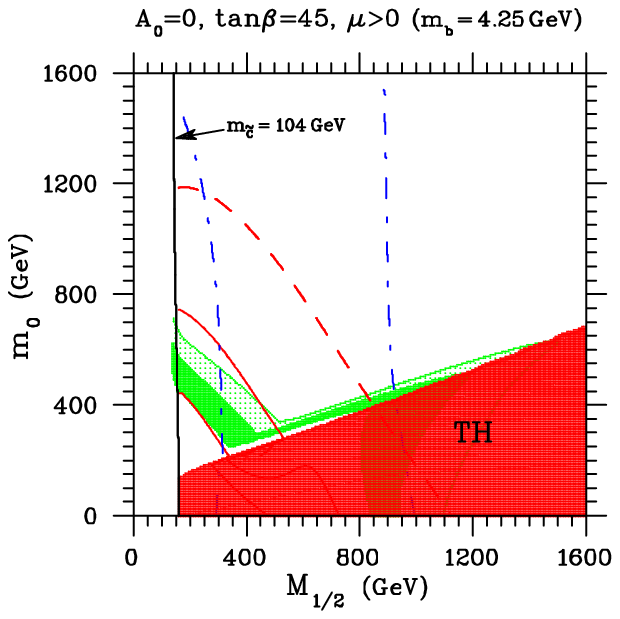} 
\vspace*{0.5cm}

\includegraphics[height=7.5cm,width=7.5cm]{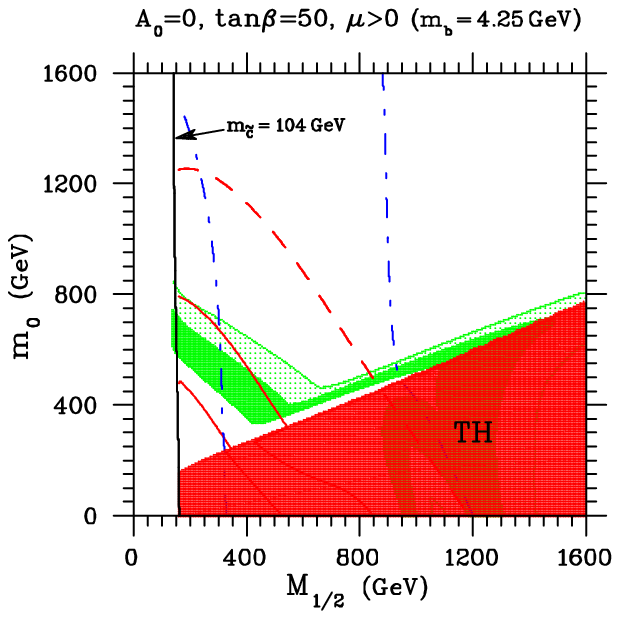}
\hspace*{.5cm}
\includegraphics[height=7.5cm,width=7.5cm]{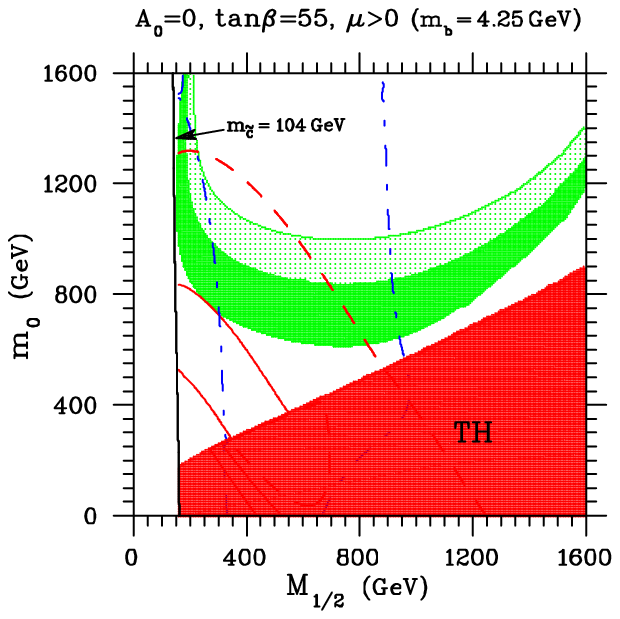}

\caption[]{
Cosmologically allowed regions of the relic density for four different values
 of $\tan \beta$ in the $(M_{1/2},m_0)$ plane. 
The remaining inputs are shown in each figure. 
The mass of the top is taken $175\GeV$. In the dark
green shaded area $0.08<\relic<0.18$. In the light green shaded
area $0.18<\relic<0.30\;$. The solid red lines mark the region within which
the supersymmetric contribution to the anomalous magnetic moment of the
muon is
$\alpha^{SUSY}_{\mu} = (43.0\pm 16.0) \times 10^{-10}$.
The dashed red line
is the boundary of the region for which the lower bound is moved to
$11.2 < 10^{10} \; \alpha^{SUSY}_{\mu}$. The dashed-dotted blue lines are
the boundaries of the region $113.5 \GeV \leq m_{Higgs} \leq 117.0 \GeV$.}

\label{fig2}  
\end{figure}

In the panels shown in figure~\ref{fig2} we display our results by drawing 
the cosmologically
allowed region $0.08<\relic<0.18$ (dark green) in the $m_0, M_{1/2}$ plane 
for values of $\tan \beta$ equal to $40$, $45$, $50$ and $55$ respectively.
Also drawn (light green) is the region $0.18<\relic<0.30$.
In the figures shown the default values for the top, tau and bottom masses
are assumed. The remaining inputs are shown on the top of each panel.
The solid red mark the region within which
the supersymmetric contribution to the anomalous magnetic moment of the
muon falls within the E821 range 
$ \alpha^{SUSY}_{\mu} = ( 43.0 \pm 16.0 ) \times 10^{-10}$.
The dashed red line
marks the boundary of the region when the more relaxed lower
bound $11.2 \times 10^{-10} \leq \alpha^{SUSY}_{\mu}$
is used \cite{Narison}, corresponding 
to the $2\sigma$ lower bound of the E821 range.
Along the blue dashed-dotted lines the light $CP$-even Higgs mass takes
values $113.5\GeV$ (left) and $117.0\GeV$ (right) respectively.
The line on the left
marks therefore the recent LEP bound on the Higgs mass \cite{LEP}.
Also shown is the chargino mass bound $104\GeV$
{\footnote {In the context of our analysis
focus point regions  \cite{focus} show up for
smaller values of the top mass. 
At this point we therefore agree with the findings of
Ref.~\cite{Ellis, ENO}. 
In any case the bulk of the focus point region 
appears  for rather large
values of $m_0$ and hence they
are not favoured by the $g - 2$ data.}}.
The shaded area (in red)
at the bottom of each figure, labelled by TH, is theoretically disallowed 
since the light stau is lighter than the lightest of the neutralinos.
From the displayed figures
we observe that for values of $\tan \beta$ up to $50$ the cosmological data 
put an upper bound on the parameter $m_0$. 
However, there is practically no such upper bound for the parameter $M_{1/2}$, 
due to the coannihilation effects \cite{Ellis} which allow for $M_{1/2} $
as large as $ 1700\GeV$ within 
the narrow coannihilation band lying above the theoretically disallowed
region.
For $\tan \beta = 55 $ a large region opens up within which the relic density
is cosmologically allowed. This is due to the pair annihilation of the
neutralinos through the pseudo-scalar Higgs exchange in the $s$-channel.
As explained before, for such high $\tan \beta$ the ratio $m_A / 2 m_{\lsp}$
approaches unity and the pseudo-scalar exchange dominates yielding
large cross sections and hence small neutralino relic densities. In this
case the lower bound put by the $g-2$ data cuts the cosmologically allowed
region which would otherwise allow for very large values of $m_0, M_{1/2}$.
The importance of these corridors has been stressed in the analysis of
\cite{Ellis} and \cite{ENO}. 
However, in our case these show up at much higher values of the
parameter $\tan \beta$. We should remark at this point that in our analysis 
we use the value of $\alpha_{strong}(M_Z)$ as input and relax unification
of the $\alpha_3$ gauge coupling with the others. For reasons already
explained, in the constrained scenario it is almost impossible to reconcile
gauge coupling unification with a value for $\alpha_{strong}(M_Z)$
consistent with experiment due to the low energy threshold effects. This
change affects drastically the values of other parameters and especially
that of the Higgsino ($\mu$) and  Higgs ($m_3^2$) mixing parameters that
in turn affect the
pseudo-scalar Higgs boson mass which plays 
a dominant role. For the $\tan \beta = 55$ case, 
close the highest possible value, and considering the conservative
lower bound on the muon's anomalous magnetic moment
$\alpha_{\mu}^{SUSY} \geq 11.2 \times 10^{-10}$
and values of
$\relic$ in the range $0.13\pm0.05$, 
we find that the point
with the highest value of $m_0$ is (in GeV) at
$(m_0, M_{1/2}) = (950, 300)$ and that with the highest value of 
$M_{1/2}$ is at $(m_0, M_{1/2}) = (600,750)$. 
The latter marks the 
lower end of
the line segment of the boundary $\alpha_{\mu}^{SUSY} = 11.2 \times 10^{-10}$ 
which amputates the cosmologically allowed stripe.
For the case displayed in the bottom right  panel of
the figure \ref{fig2} 
the upper mass limits put on the LSP, and the lightest
of the charginos, stops and the staus are
$m_{\lsp} < 287, m_{{\tilde \chi}^{+}} <  539,  m_{\tilde t} < 1161,
m_{\tilde \tau} < 621$ (in $\GeV$). 
Allowing for $A_0 \neq 0$ values, the upper bounds put on $m_0, M_{1/2}$
increase a little and so do the aforementioned bounds on the sparticle
masses.
Thus it appears that the prospects of discovering CMSSM at  a
$e^{+} e^{-}$ collider with center of mass energy $\sqrt s = 800 \GeV$,
such as TESLA, are not guaranteed. 
However in the allowed regions 
the next to the lightest neutralino, ${\tilde{\chi}^{\prime}}$, has a mass
very close to the lightest of the charginos and hence the process
$e^{+} e^{-} \goes {\tilde{\chi}} {\tilde{\chi}^{\prime}}$,
with 
${\tilde{\chi}^{\prime}}$ subsequently decaying to
$ {\tilde{\chi}} + {l^{+}} {l^{-}}$ or 
$ {\tilde{\chi}}+\mathrm{2\,jets}$,  
is kinematically allowed for such large $\tan \beta$, provided
the energy is increased to at least $\sqrt{s} = 900 \GeV$. It should
be noted however that this channel proceeds via the $t$-channel exchange
of a selectron  is suppressed due to the heaviness of the exchanged
sfermion.

The situation changes, however, when the strict E821 limits are imposed 
$\alpha_{\mu}^{SUSY} = (43.0 \pm 16.0) \times 10^{-10}$. For instance 
in the $\tan \beta = 55$ case displayed in figure \ref{fig2} there is no  
cosmologically allowed region which obeys this bound.
For the other cases, $\tan \beta < 50$, 
the maximum allowed $M_{1/2}$ is about $475\GeV$, occurring at
$m_0 \simeq 375\GeV$, and the maximum $m_0$ is $ 600\GeV$ when
$M_{1/2} \simeq 300\GeV$. The upper limits on the masses of the sparticles
quoted previously reduce to 
$m_{\lsp} < 192, m_{{\tilde \chi}^{+}}  <  353,  m_{\tilde t} < 775,
m_{\tilde \tau} < 436$ all in GeV.
However, these values 
refer to the limiting case $A_0 = 0$. Scanning the parameter space 
allowing also for $A_0 \neq 0$ we obtain the upper limits displayed in
the table~\ref{table1}. In this the unbracketed values correspond to the E821
limits on the $g-2$. For completeness we also display, within brackets, the
bounds obtained when the weaker lower bound
$\alpha_{\mu}^{SUSY} \geq 11.2 \times 10^{-10}$ is imposed.
We see that even at TESLA with
center of mass energy $\sqrt s = 800\GeV$, the prospects of discovering
CMSSM are guaranteed in the
$e^{+} e^{-} \goes {\tilde \chi}^{+} {{\tilde \chi}^{-}} $
if the E821 bounds are imposed.

\begin{figure}[t] 
\centering\includegraphics[height=10cm,width=10cm]{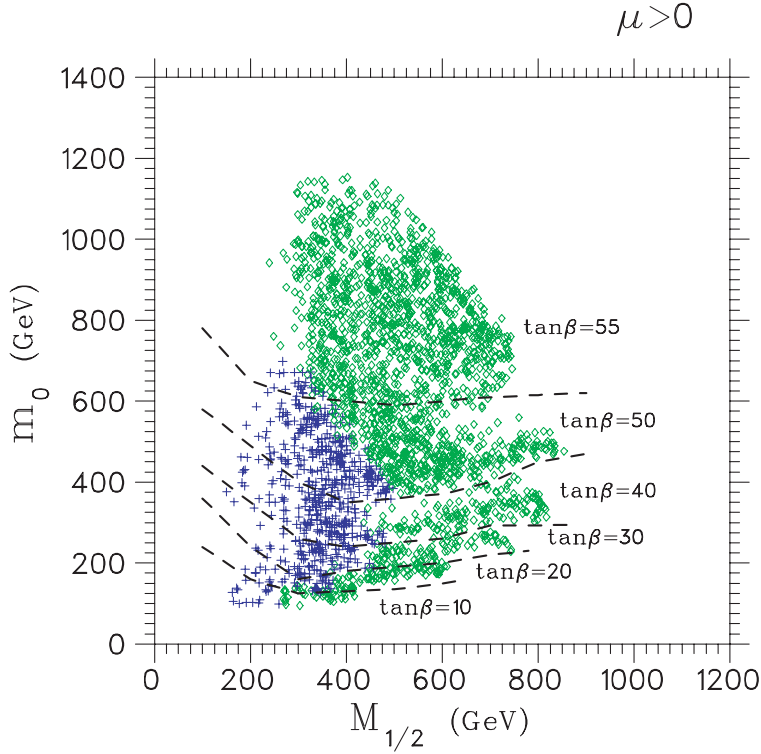}
\caption[]{
In  the $(M_{1/2},m_0)$ plane, we display all
points compatible with
$\alpha_{\mu}^{SUSY} = ( 43.0 \pm 16.0 ) \times 10^{-10}$ ($+$) and 
$11.2 \times 10^{-10}<\alpha_{\mu}^{SUSY}<59.0 \times 10^{-10}$
($\diamond$).
All the points are consistent with 
the cosmological bound $\relic = 0.13\pm 0.05$
and they  are grouped in regions, separated by dashed contours 
each of which is the boundary of $\tan \beta $ with the value shown
beneath. 
In the top region, designated by $\tan \beta = 55$,
the parameter $\tan \beta $ takes values between 50 and 55.
}
\label{fig3}  
\end{figure}

In the figure \ref{fig3} we display in the $(M_{1/2},m_0)$ plane
the points which are consistent both with the muon's anomalous magnetic
moment bounds mentioned before and cosmology, as well as with the other
accelerators data.
Each of the points is taken from a sample of 40,000 random points in the 
part of the parameter
space  defined by $m_0 < 1.5 \TeV$, $M_{1/2} < 1.5 \TeV$, 
$|A_0| < 1 \TeV$ and $2<\tan\beta<55$.
All the points are consistent with 
the cosmological bound $\relic = 0.13\pm 0.05$.
The  plus points (colored in blue) are those consistent with the E821 bound  
$ 27 \times 10^{-10} < \alpha_{\mu}^{SUSY} < 59 \times 10^{-10} $,
while the
diamonds (colored in green) are consistent with the  more relaxed  bound
$11.2 \times 10^{-10}< \alpha_{\mu}^{SUSY} <59 \times 10^{-10}$.
The points are grouped in regions, separated by dashed contours,
each of which constitutes  
the boundary of $\tan \beta$ with the value shown beneath. 
In the region designated as $\tan \beta = 55$ all points have
$55 > \tan \beta > 50$. 
It is seen clearly that only a few points in the $\tan \beta > 50$ case
can survive the E821 bound. For $\tan \beta < 50$  the parameter
$M_{1/2}$ cannot be larger than about $500\GeV$, attaining its maximum
value at $m_0 \simeq 400\GeV$ , and the maximum $m_0$ is $725 \GeV$ 
occurring at  $M_{1/2}\simeq 275\GeV$.
The upper limits put on $m_0, M_{1/2}$ 
result to the sparticle mass bounds displayed in the
table~{\ref{table1}}.

\section{Conclusions}
In this work 
we have undertaken the problem of calculating the neutralino relic density
in the framework of the CMSSM, by paying special attention to the
pseudo-scalar 
Higgs exchange mechanism which is dominant in the large $\tan \beta$ region.
Imposing the bounds  
$ \alpha^{SUSY}_{\mu} = ( 43.0 \pm 16.0 ) \times 10^{-10}$ 
on the muon's anomalous magnetic moment, put by the BNL E821 experiment, 
in combination with the cosmological data $\relic = 0.13 \pm 0.05$, 
severely restricts the sparticle spectrum.
We found that 
the pseudo-scalar Higgs exchange mechanism opens cosmologically allowed
corridors, of high $m_0, M_{1/2}$, only for very large $\tan \beta > 50$, 
which, however, have little overlap with the regions allowed by the E821 data. 
In fact  
only a few isolated points in the parameter space with $\tan \beta > 50$
can survive the restrictions imposed by both data.
The bounds put on the sparticle spectrum can guarantee that in LHC  
but also in 
a $e^{+}e^{-}$ linear collider with center of mass energy
$\sqrt{s} = 800\GeV$, such as TESLA, CMSSM can be discovered.
The guarantee for a TESLA machine with this energy is
lost in a charged sparticle final state channel, 
if the lower bound on the value of $g -2 $ is lowered to its
$\approx 2 \sigma$ value, but not for the LHC. 
In this case only by increasing the
center of mass energy to be $\simeq 1.2 \TeV$, a
$e^{+}e^{-}$ linear collider can find CMSSM in 
$\tilde{\tau} \, {\tilde{\tau}}^*$ or
${\tilde \chi}^{+} {{\tilde \chi}^{-}} $ channels.

\vspace{1.8cm}
\noindent 
{\bf Acknowledgements} \\ 
\noindent 
A.B.L. acknowledges support from HPRN-CT-2000-00148
and HPRN-CT-2000-00149 programmes. He also thanks the University of Athens
Research Committee for partially supporting this work. 
V.C.S. acknowledges support by a Marie Curie Fellowship of the EU
programme IHP under contract HPMFCT-2000-00675.
The authors thank D.V.~Nanopoulos,  H.~Eberl, 
S.~Kraml and W.~Majerotto for valuable discussions.


\clearpage
\begin{table}[b]
\begin{center}
\begin{tabular}{|c|c|c|c|c|c|} \hline \hline
$\tan\beta$ & $\lsp^0$ & $\tilde{\chi}^+$ & $\tilde{\tau}$ & $\tilde{t}$ & $h$ 
                                                                    \\ \hline
  10  &  108 (174) & 184 (306) & 132 (197) & 376 (686) &  115 (116) \\
  20  &  154 (255) & 268 (457) & 175 (274) & 603 (990) &  116 (118) \\
  30  &  191 (310) & 338 (560) & 212 (312) & 740 (1200) &  117 (118) \\
  40  &  201 (340) & 357 (617) & 274 (353) & 785 (1314) &  117 (119) \\
  50  &  208 (357) & 371 (646) & 440 (427) & 822 (1357) &  117 (119) \\
  55  &  146 (311) & 260 (563) & 424 (676) & 606 (1237) &  115 (117) \\ 
 \hline \hline
\end{tabular}
\end{center}

\vspace{.4cm}
\caption{Upper bounds, in GeV, 
on the masses of the lightest of the neutralinos,
charginos, staus, stops and  Higgs bosons for various values of
$\tan\beta$ if the the E821 bounds are imposed.
The values within brackets represent the same situation when the weaker
bounds
$11.2 \times 10^{-10}<\alpha_{\mu}^{SUSY}<59.0 \times 10^{-10}$
are used (see main text).}

\label{table1}
\end{table}

\clearpage

\end{document}